\documentstyle[12pt]{article}
\newlength{\extraspace}
\newlength{\extraspaces}
\newcommand{\be}{\begin{equation}
\addtolength{\abovedisplayskip}{\extraspaces}
\addtolength{\belowdisplayskip}{\extraspaces}
\addtolength{\abovedisplayshortskip}{\extraspace}
\addtolength{\belowdisplayshortskip}{\extraspace}}
\newcommand{\ee}{\end{equation}}
\newcommand{\ba}{\begin{eqnarray}
\addtolength{\abovedisplayskip}{\extraspaces}
\addtolength{\belowdisplayskip}{\extraspaces}
\addtolength{\abovedisplayshortskip}{\extraspace}
\addtolength{\belowdisplayshortskip}{\extraspace}}
\newcommand{\ea}{\end{eqnarray}}
\makeatletter

\@addtoreset{equation}{section}
\makeatother
\newcommand{\nonu}{\nonumber \\[.5mm]}
\newcommand{\A}{&\!\!\!}
\begin{document}
\thispagestyle{empty}
\setlength{\baselineskip}{6mm}
%
%
%
\vspace*{7mm}
\begin{center}
{\large{\bf Physical significances of NL/L SUSY relation}} \\[20mm]
{\sc Kazunari Shima}
\footnote{
\tt e-mail: shima@sit.ac.jp} \ 
and \ 
{\sc Motomu Tsuda}
\footnote{
\tt e-mail: tsuda@sit.ac.jp} 
\\[5mm]
{\it Laboratory of Physics, 
Saitama Institute of Technology \\
Fukaya, Saitama 369-0293, Japan} \\[20mm]
\begin{abstract}
We discuss explicitly the details and some remarkable physical consequences 
of the NL/L SUSY relation between $N = 2$ LSUSY QED and $N = 2$ NLSUSY 
in two-dimensional space-time, 
which may show the potential of the SUSY composite model 
for the theory of everything beyond the SM.  \\[5mm]
\end{abstract}
\end{center}

\newpage

\section{Introduction}

\noindent
SUSY \cite{WZ,VAa} and its spontaneous breakdown are profound notions related to the space-time symmetry 
and play crucial roles for the unified description of space-time and matter, threrefore, 
to  be studied  not only in the low energy particle physics 
but also in the cosmology, i.e. in the framework necessarily accomodating graviton.    

The nonlinear supersymmmetric general relativity (NLSUSY GR) theory \cite{KS1} is constructed 
based upon the nonlinear (NL) representation of SUSY \cite{VAa,VS} and the principle of 
the general relativity (GR) for four-dimensional space-time (d = 4) manifold 
equipped with NLSUSY degrees of freedom (d.o.f.) on tangent space, 
which proposes a new paradigm (called SGM scenario \cite{KS1,ST1,ST2}) 
for describing the unity of nature. 
The NLSUSY invariant GR action $L_{\rm NLSUSYGR}(w)$ (in terms of the unified vierbein $w^a{}_\mu$) 
desribes geometrically the basic principle, 
i.e. the ultimate shape of nature is unstable space-time with the constant energy density 
(cosmological term) $\Lambda > 0$. 
NLSUSY GR decays (called {\it Big Dcay}) spontaneously and creates  quantum mechanically  
ordinary Riemann space-time and spin-${1 \over 2}$ massless Nambu-Goldstone (NG) fermion matter
(called {\it superon})  with the potential (cosmological term) $V = \Lambda > 0$ depicted by the SGM action 
$L_{\rm SGM}(e, \psi)$ (in terms of the ordinary vierbein $e^a{}_\mu$ and the NG fermions $\psi^i$),
which ignites the Big Bang of the universe. 
We showed explicitly in our previous works 
in asymptotic Riemann-flat ($e^a{}_\mu \rightarrow \delta^a{}_\mu$) space-time 
that the vacuum (true minimum) in SGM scenario is $V = 0$ 
{\it which is achieved when the NG fermions constitute the linear (L) SUSY representation} 
dictated by the space-time $N$-extended NLSUSY symmetry. 
That is, all local fields of the LSUSY supermultiplet and the familiar LSUSY gauge theory 
$L_{\rm LSUSY}(A, \lambda, v_a, \dots)$ emerge as the composites of NG fermions 
on the true vacuum. 
In the SGM scenario all (observed) particles are assigned uniquely 
into a single irreducible representation of $SO(N)$ ($SO(10)$) super-Poincar\'e (SP) group 
as an on-shell supermultiplet of $N$ LSUSY \cite{KS2}. 
As mentioned above, they are considered to be realized 
as (massless) eigenstates of $SO(N)$ ($SO(10)$) SP composed of NG fermions (superons) 
which correspond to the coset space coordinates 
for $super-GL(4,R) \over GL(4,R)$ of NLSUSY describing the spontaneous SUSY breaking (SSB) by itself. 

In order to extract the low energy physical contents of NLSUSY GR and to examine the SGM scenario, 
we investigated the NLSUSY GR model in asymptotic Riemann-flat space-time, 
where the NLSUSY model appears from the cosmological term of NLSUSY GR. 
The NLSUSY model \cite{VAa} are recasted (related) systematically to various LSUSY ({\it free}) theories 
with the SSB \cite{IK1}-\cite{lin-ST2} 
and also to {\it interacting} $N = 2$ LSUSY (Yukawa-interaction and QED) theories 
in two-dimensional space-time ($d = 2$) \cite{lin-ST3}-\cite{lin-ST4b} 
under the adoption of the simplest {\it SUSY invariant constraints} 
and the subsequent {\it SUSY invariant relations} in {\it NL/L SUSY relation}. 
(Note that $N = 2$ SUSY in SGM scenario realizes $J^P = 1^-$ vector field \cite{STT2}. 
Therefore $N = 2$ SUSY is the viable minimal case.) 
The SUSY invariant relations, which are obtained {\it systematically} from the SUSY invariant constraints 
\cite{IK1,UZ} in the superfield formulation \cite{WB}, describe all component fields 
in the LSUSY multiplets as the composites of the NG-fermion superons 
and give the relations between the NLSUSY model and the LSUSY theories with the SSB. 
Through the NL/L SUSY relation, we pointed out that 
the scale of the SSB in NLSUSY which is naturally related to the cosmological term of NLSUSY GR 
gives a simple explanation of the mysterious (observed) numerical relation 
between the (four-dimensional) dark energy density of the universe and the neutrino mass \cite{ST2,ST5} 
in the vacuum of the $N = 2$ SUSY QED theory \cite{STL}. 

In the above SGM scenario it is an interesting and crucial problem 
to study how the (bare) gauge coupling constant of the SUSY QED theory 
is determined or whether it is calculable or not, provided the theory is that of everything. 
In this paper we study the $N = 2$ SUSY QED theory in $d = 2$ 
for simplicity of calculations without loss of generality by linearizing $N = 2$ NLSUSY 
under {\it general} SUSY invariant constraints which produce the subsequent general SUSY invariant relations. 
The NL/L SUSY relation for the SUSY QED theory are discussed generally 
as self-contained as possible. 
%
%
We find in the general NL/L SUSY relation a remarkable mechanism 
which determines the magnitude of the bare gauge coupling constant 
from constant terms in SUSY invariant relations 
(vacuum expectation values (v.e.v.'s)) of auxiliary fields. 
We show that the NL/L SUSY relation determines the magnitude of 
the electromagnetic coupling constant (i.e. the fine structure constant) of $N = 2$ LSUSY QED 
%
and argue remarkable physical significances of the NL/L SUSY relation. 
\section{$N = 2$ LSUSY QED action }

\noindent

We construct manifestly $N = 2$ LSUSY QED action  in $d = 2$.
Let us first introduce the superfield formulation of $N = 2$ SUSY QED theory in $d = 2$, 
in which a $N = 2$ (general) SUSY QED action is constructed from $N = 2$ general gauge 
and $N = 2$ scalar (matter) superfields on superspace coordinates $(x^a, \theta_\alpha^i)$ ($i = 1, 2$). 
The $d = 2$, $N = 2$ general gauge superfield \cite{DVF,ST4} is defined by 
\ba
{\cal V}(x, \theta) \A = \A C(x) + \bar\theta^i \Lambda^i(x) 
+ {1 \over 2} \bar\theta^i \theta^j M^{ij}(x) 
- {1 \over 2} \bar\theta^i \theta^i M^{jj}(x) 
+ {1 \over 4} \epsilon^{ij} \bar\theta^i \gamma_5 \theta^j \phi(x) 
\nonu
\A \A 
- {i \over 4} \epsilon^{ij} \bar\theta^i \gamma_a \theta^j v^a(x) 
- {1 \over 2} \bar\theta^i \theta^i \bar\theta^j \lambda^j(x) 
- {1 \over 8} \bar\theta^i \theta^i \bar\theta^j \theta^j D(x), 
\label{VSF}
\ea
where component fields $\varphi_{\cal V}^I(x) = \{ C(x), \Lambda^i(x), M^{ij}(x), \cdots \}$ 
in the superfield (\ref{VSF}) are denoted 
by $(C, D)$ for two scalar fields, $(\Lambda^i, \lambda^i)$ for two doublet (Majorana) spinor fields, 
$\phi$ for a pseudo scalar field, $v^a$ for a vector field, 
and $M^{ij} = M^{(ij)}$ $\left(= {1 \over 2}(M^{ij} + M^{ji}) \right)$ 
for three scalar fields ($M^{ii} = \delta^{ij} M^{ij}$). 
The $d = 2$, $N = 2$ scalar superfield is expressed as 
\ba
\Phi^i(x, \theta) \A = \A B^i(x) + \bar\theta^i \chi(x) - \epsilon^{ij} \bar\theta^j \nu(x) 
- {1 \over 2} \bar\theta^j \theta^j F^i(x) + \bar\theta^i \theta^j F^j(x) 
- i \bar\theta^i \!\!\not\!\partial B^j(x) \theta^j 
\nonu
\A \A 
+ {i \over 2} \bar\theta^j \theta^j (\bar\theta^i \!\!\not\!\partial \chi(x) 
- \epsilon^{ik} \bar\theta^k \!\!\not\!\partial \nu(x)) 
+ {1 \over 8} \bar\theta^j \theta^j \bar\theta^k \theta^k \Box B^i(x), 
\label{SSF}
\ea
where component fields $\varphi_\Phi^I(x) = \{ B^i(x), \chi^i(x), \nu^i(x), F^i(x) \}$ 
in the superfield (\ref{SSF}) are denoted by $B^i$ for doublet scalar fields, 
$(\chi, \nu)$ for two (Majorana) spinor fields 
and $F^i$ for doublet auxiliary scalar fields. 
The supertransformations of the gauge and scalar superfields with constant (Majorana) spinor parameters $\zeta^i$ 
are given as 
\be
\delta_\zeta {\cal V}(x, \theta) = \bar\zeta^i Q^i {\cal V}(x, \theta), 
\ \ \delta_\zeta \Phi^i(x, \theta) = \bar\zeta^j Q^j \Phi^i(x, \theta), 
\label{SFtransfn}
\ee
where supercharges $Q^i = {\partial \over \partial\bar\theta^i} + i \!\!\not\!\partial \theta^i$, 
which determine LSUSY transformations of the component fields 
in the power series expansion with respect to $\theta^i$. 

The general $N = 2$ SUSY QED (gauge) action (with SSB) is written in terms of the general gauge 
and the scalar (matter) superfields (\ref{VSF}) and (\ref{SSF}) (for the massless case) as 
\be
L^{\rm gen.}_{N = 2{\rm SUSYQED}} 
= L_{{\cal V}{\rm kin}} + L_{{\cal V}{\rm FI}} 
+ L_{\Phi{\rm kin}} + L_e 
\label{SQEDaction}
\ee
with 
\ba
L_{{\cal V}{\rm kin}} 
\A = \A {1 \over 32} \left\{ \int d^2 \theta^i 
\ (\overline{D^i {\cal W}^{jk}} D^i {\cal W}^{jk} 
+ \overline{D^i {\cal W}_5^{jk}} D^i {\cal W}_5^{jk}) \right\}_{\theta^i = 0}, 
\label{Vkin}
\\
L_{{\cal V}{\rm FI}} 
\A = \A {1 \over 2} \int d^4 \theta^i \ {\xi \over \kappa} {\cal V}, 
\label{VFI}
\\
L_{\Phi{\rm kin}} + L_e 
\A = \A - {1 \over 16} \int d^4 \theta^i \ e^{-4e{\cal V}} (\Phi^j)^2, 
\label{gauge}
\ea
where $L_{{\cal V}{\rm kin}}$, $L_{{\cal V}{\rm FI}}$, $L_{\Phi{\rm kin}}$ and $L_e$ 
are the kinetic terms for the vector supermultiplet, the Fayet-Iliopoulos (FI) $D$ term, 
the kinetic terms for the scalar (matter) supermultiplet and the gauge interaction terms, respectively. 
In Eq.(\ref{Vkin}) ${\cal W}^{ij}$ and ${\cal W}_5^{ij}$ are 
scalar and pseudo scalar superfields defined by 
\be
{\cal W}^{ij} = \bar D^i D^j {\cal V}, \ \ \ {\cal W}_5^{ij} = \bar D^i \gamma_5 D^j {\cal V} 
\ee
with the differential operators $D^i = {\partial \over \partial\bar\theta^i} - i \!\!\not\!\!\partial \theta^i$. 
In Eq.(\ref{VFI}) $\xi$ is an arbitrary dimensionless parameter 
and $\kappa$ is a constant with the dimension (mass)$^{-1}$, 
while in Eq.(\ref{gauge}) $e$ is a gauge coupling constant whose dimension is (mass)$^1$ in $d = 2$. 

The explicit component form of the $N = 2$ SUSY QED action (\ref{SQEDaction}), 
i.e. the actions from (\ref{Vkin}) to (\ref{gauge}), is 
\ba
L_{{\cal V}{\rm kin}} \A = \A 
- {1 \over 4} (F_{0ab})^2 
+ {i \over 2} \bar\lambda_0^i \!\!\not\!\partial \lambda_0^i 
+ {1 \over 2} (\partial_a A_0)^2 + {1 \over 2} (\partial_a \phi_0)^2 + {1 \over 2} D_0^2 
\equiv L^0_{{\cal V}{\rm kin}}, 
\label{Vkin-comp}
\\
L_{{\cal V}{\rm FI}} \A = \A - {\xi \over \kappa} (D_0 - \Box C) 
\equiv L^0_{{\cal V}{\rm FI}} + {\xi \over \kappa} \Box C, 
\label{VFI-comp}
\\
L_{\Phi{\rm kin}} \A = \A 
{i \over 2} \bar\chi \!\!\not\!\partial \chi 
+ {1 \over 2} (\partial_a B^i)^2 
+ {i \over 2} \bar\nu \!\!\not\!\partial \nu 
+ {1 \over 2} (F^i)^2 
- {1 \over 4} \partial_a (B^i \partial^a B^i) 
\nonu
\A \equiv \A L^0_{\Phi{\rm kin}} - {1 \over 4} \partial_a (B^i \partial^a B^i), 
\label{Skin-comp}
\\
L_e \A = \A 
e \ \bigg\{ i v_{0a} \bar\chi \gamma^a \nu 
- \epsilon^{ij} v_0^a B^i \partial_a B^j 
+ \bar\lambda_0^i \chi B^i 
+ \epsilon^{ij} \bar\lambda_0^i \nu B^j 
- {1 \over 2} D_0 (B^i)^2 
\nonu
\A \A 
+ {1 \over 2} A_0 (\bar\chi \chi + \bar\nu \nu) 
- \phi_0 \bar\chi \gamma_5 \nu + \cdots \bigg\} 
\nonu
\A \A 
+ {1 \over 2} e^2 \{ (v_{0a}{}^2 - A_0^2 - \phi_0^2) (B^i)^2 + \cdots \} 
+ \cdots, 
\nonu
\A \equiv \A L^0_e + \cdots, 
\label{gauge-comp}
\ea
where we use gauge invariant quantities \cite{lin-ST4a,WB} denoted by 
\be
\{ A_0, \phi_0, F_{0ab}, \lambda_0^i, D_0 \} 
\equiv \{ M^{ii}, \phi, F_{ab}, \lambda^i + i \!\!\not\!\partial \Lambda^i, D + \Box C \} 
\label{gauge-inv}
\ee
with $F_{0ab} = \partial_a v_{0b} - \partial_b v_{0a}$ and $F_{ab} = \partial_a v_b - \partial_b v_a$, 
which are invariant ($v_{0a} = v_a$ transforms as an Abelian gauge field) 
under a SUSY generalized gauge transformation, $\delta_g {\cal V} = \Lambda^1 + \Lambda^2$, 
with generalized gauge parameters $\Lambda^i$ 
in the form of the $N = 2$ scalar superfields. 
In Eqs. from (\ref{Vkin-comp}) to (\ref{gauge-comp}) $L^0_{{\cal V}{\rm kin}}$, $L^0_{{\cal V}{\rm FI}}$, 
$L^0_{\Phi{\rm kin}}$ and $L^0_e$ are defined as the actions which are expressed 
in terms of only the component fields $\{ A_0, \phi_0, v_{0a}, \lambda_0^i, D_0 \}$ 
corresponding to the d.o.f. for the minimal off-shell vector supermultiplet 
and the components for the scalar supermultiplet, 
while the ellipses in Eq.(\ref{gauge-comp}) mean the terms depending on the redundant auxiliary fields 
$\{ C, \Lambda, M^{ij} (i \not= j) \}$ in the general gauge superfield 
and higher order terms of $e^n \ (n \ge 3)$. 
The $N = 2$ SUSY QED action (\ref{SQEDaction}) in the Wess-Zumino (WZ) gauge \cite{WB,ST4} gives 
the minimal action for the minimal off-shell vector supermultiplet with the arbitrary $e$.

\section{Linearization of $N = 2$ NLSUSY}

In order to discuss the (general) NL/L SUSY relation for the $N = 2$ SUSY QED theory in $d = 2$ 
without imposing a priori any special gauge conditions, 
let us show explicitly the SUSY invariant constraints and the subsequent SUSY invariant relations 
in the most general form. 
They are obtained by introducing NG fermions $\psi^i$ with their supertransformations 
(on a hypersurface defined by $\theta^i = \kappa \psi^i$ in the superspace), 
i.e. NLSUSY transformations \cite{VAa}, 
\be
\delta_\zeta \psi^i = {1 \over \kappa} \zeta^i 
- i \kappa \bar\zeta^j \gamma^a \psi^j \partial_a \psi^i. 
\label{NLSUSY}
\ee
The $N = 2$ NLSUSY transformations (\ref{NLSUSY}) make the following action invariant; 
namely, the $N = 2$ NLSUSY action is 
\be
L_{N = 2{\rm NLSUSY}} = - {1 \over {2 \kappa^2}} \ \vert w \vert, 
\label{NLSUSYaction}
\ee
where $\vert w \vert$ is the determinant \cite{VAa} describing the dynamics of (massless) $\psi^i$, 
i.e. in $d = 2$, 
\be
\vert w \vert = \det(w^a{}_b) = \det(\delta^a_b + t^a{}_b) 
= 1 + t^a{}_a + {1 \over 2!}(t^a{}_a t^b{}_b - t^a{}_b t^b{}_a) 
\label{det-w}
\ee
with $t^a{}_b = - i \kappa^2 \bar\psi^i \gamma^a \partial_b \psi^i$. 

The SUSY invariant relations, which describe all the component fields 
in the $N = 2$ SUSY QED theory as the composites of the NG fermions $\psi^i$, 
are systematically obtained by considering the superfields on specific supertranslations 
of superspace coordinates \cite{IK1,UZ} with a parameter $\zeta^i = - \kappa \psi^i$, 
which are denoted by $(x'^a, \theta_\alpha'^i)$, 
\ba
\A \A 
x'^a = x^a + i \kappa \bar\theta^i \gamma^a \psi^i, 
\nonu
\A \A 
\theta'^i = \theta^i - \kappa \psi^i, 
\ea
where the origin of $(x'^a, \theta_\alpha'^i)$ 
corresponds to the hypersurface $\theta^i = \kappa \psi^i$ for NLSUSY. 
Indeed, we define the $N = 2$ general gauge and the $N = 2$ scalar (matter) superfields 
on $(x'^a, \theta_\alpha'^i)$ as 
\be
{\cal V}(x', \theta') \equiv \tilde{\cal V}(x, \theta), 
\ \ \Phi^i(x', \theta') \equiv \tilde \Phi^i(x, \theta), 
\label{SFpsi}
\ee
where $\tilde{\cal V}(x, \theta)$ and $\tilde \Phi^i(x, \theta)$ 
may be expanded as 
\ba
\tilde{\cal V}(x, \theta) 
\A = \A \tilde C(x) + \bar\theta^i \tilde\Lambda^i(x) 
+ {1 \over 2} \bar\theta^i \theta^j \tilde M^{ij}(x) 
- {1 \over 2} \bar\theta^i \theta^i \tilde M^{jj}(x) 
+ {1 \over 4} \epsilon^{ij} \bar\theta^i \gamma_5 \theta^j \tilde\phi(x) 
\nonu
\A \A 
- {i \over 4} \epsilon^{ij} \bar\theta^i \gamma_a \theta^j \tilde v^a(x) 
- {1 \over 2} \bar\theta^i \theta^i \bar\theta^j \tilde\lambda^j(x) 
- {1 \over 8} \bar\theta^i \theta^i \bar\theta^j \theta^j \tilde D(x), 
\label{tVSF} \\
\tilde \Phi^i(x, \theta) 
\A = \A \tilde B^i(x) + \bar\theta^i \tilde \chi(x) - \epsilon^{ij} \bar\theta^j \tilde \nu(x) 
- {1 \over 2} \bar\theta^j \theta^j \tilde F^i(x) + \bar\theta^i \theta^j \tilde F^j(x) + \cdots. 
\label{tSSF}
\ea
In Eqs.(\ref{tVSF}) and (\ref{tSSF}) the component fields 
$\tilde\varphi_{\cal V}^I(x) = \{ \tilde C(x), \tilde\Lambda^i(x), \cdots \}$ 
and $\tilde\varphi_\Phi^I(x) = \{ \tilde B^i(x), \tilde\chi(x), \cdots \}$ 
can be expressed in terms of the initial component fields $\varphi_{\cal V}^I(x)$ and $\varphi_\Phi^I(x)$ 
in Eqs.(\ref{VSF}) and (\ref{SSF}) and the NG fermions $\psi^i$ \cite{lin-ST2,lin-ST4b}. 
According to the supertransformations (\ref{SFtransfn}) and (\ref{NLSUSY}), 
the superfields (\ref{SFpsi}) transform homogeneously \cite{IK1,UZ} as 
\be
\delta_\zeta \tilde{\cal V}(x, \theta) = \xi^a \partial_a \tilde{\cal V}(x, \theta), 
\ \ \delta_\zeta \tilde \Phi^i(x, \theta) = \xi^a \partial_a \tilde \Phi^i(x, \theta) 
\ee
with $\xi^a = i \kappa \bar\psi^i \gamma^a \zeta^i$, 
which mean the components $\tilde\varphi_{\cal V}^I(x)$ 
and $\tilde\varphi_\Phi^I(x)$ do not transform each other, respectively. 
Therefore, the following conditions, i.e. the SUSY invariant constraints eliminating the other d.o.f. 
than $\varphi_{\cal V}^I(x)$, $\varphi_\Phi^I(x)$ and $\psi^i$, can be imposed, 
\ba
\A \A 
\tilde\varphi_{\cal V}^I(x) = {\rm constant}, 
\label{SUSYconst-VSF}
\\
\A \A 
\tilde\varphi_\Phi^I(x) = {\rm constant}, 
\label{SUSYconst-SSF}
\ea
which are invariant (conserved quantities) under the supertrasformations (\ref{SFtransfn}) and (\ref{NLSUSY}). 

The constraints (\ref{SUSYconst-VSF}) and (\ref{SUSYconst-SSF}) 
are written in the most general form as follows; 
\ba
\A \A 
\tilde C = \xi_c, \ \ \tilde\Lambda^i = \xi_\Lambda^i, 
\ \ \tilde M^{ij} = \xi_M^{ij}, \ \ \tilde\phi = \xi_\phi, 
\ \ \tilde v^a = \xi_v^a, \ \ \tilde\lambda^i = \xi_\lambda^i, 
\ \ \tilde D = {\xi \over \kappa}, 
\label{SUSYconst-VSF1}
\\
\A \A 
\tilde B^i = \xi_B^i, \ \ \tilde\chi = \xi_\chi, \ \ \tilde\nu = \xi_\nu, 
\ \ \tilde F^i = {\xi^i \over \kappa}, 
\label{SUSYconst-SSF1}
\ea
where the mass dimensions of constants (or constant spinors) in $d = 2$ 
are defined by ($-1$, ${1 \over 2}$, $0$, $0$, $0$, $-{1 \over 2}$) 
for ($\xi_c$, $\xi_\Lambda^i$, $\xi_M^{ij}$, $\xi_\phi$, $\xi_v^a$, $\xi_\lambda^i$), 
($0$, $-{1 \over 2}$, $-{1 \over 2}$) for ($\xi_B^i$, $\xi_\chi$, $\xi_\nu$) and $0$ for $\xi^i$ for convenience. 
The general SUSY invariant constraints (\ref{SUSYconst-VSF1}) and (\ref{SUSYconst-SSF1}) 
can be solved with respect to the initial component fields $\varphi_{\cal V}^I$ and $\varphi_\Phi^I$ 
in terms of $\psi^i$; namely, the SUSY invariant relations $\varphi_{\cal V}^I = \varphi_{\cal V}^I(\psi)$ 
are calculated systematically and straightforwardly as 
\ba
C \A = \A \xi_c + \kappa \bar\psi^i \xi_\Lambda^i 
+ {1 \over 2} \kappa^2 (\xi_M^{ij} \bar\psi^i \psi^j - \xi_M^{ii} \bar\psi^j \psi^j) 
+ {1 \over 4} \xi_\phi \kappa^2 \epsilon^{ij} \bar\psi^i \gamma_5 \psi^j 
- {i \over 4} \xi_v^a \kappa^2 \epsilon^{ij} \bar\psi^i \gamma_a \psi^j 
\nonu
\A \A 
- {1 \over 2} \kappa^3 \bar\psi^i \psi^i \bar\psi^j \xi_\lambda^j 
- {1 \over 8} \xi \kappa^3 \bar\psi^i \psi^i \bar\psi^j \psi^j, 
\nonu
\Lambda^i \A = \A \xi_\Lambda^i 
+ \kappa (\xi_M^{ij} \psi^j - \xi_M^{jj} \psi^i) 
+ {1 \over 2} \xi_\phi \kappa \epsilon^{ij} \gamma_5 \psi^j 
- {i \over 2} \xi_v^a \kappa \epsilon^{ij} \gamma_a \psi^j 
\nonu
\A \A 
- {1 \over 2} \xi_\lambda^i \kappa^2 \bar\psi^j \psi^j 
+ {1 \over 2} \kappa^2 
(\psi^j \bar\psi^i \xi_\lambda^j 
- \gamma_5 \psi^j \bar\psi^i \gamma_5 \xi_\lambda^j 
- \gamma_a \psi^j \bar\psi^i \gamma^a \xi_\lambda^j) 
\nonu
\A \A 
- {1 \over 2} \xi \kappa^2 \psi^i \bar\psi^j \psi^j 
- i \kappa \!\!\not\!\partial C(\psi) \psi^i, 
\nonu
M^{ij} \A = \A \xi_M^{ij} 
+ \kappa \bar\psi^{(i} \xi_\lambda^{j)} 
+{1 \over 2} \xi \kappa \bar\psi^i \psi^j 
+ i \kappa \epsilon^{(i \vert k \vert} \epsilon^{j)l} \bar\psi^k \!\!\not\!\partial \Lambda^l(\psi) 
- {1 \over 2} \kappa^2 \epsilon^{ik} \epsilon^{jl} \bar\psi^k \psi^l \Box C(\psi), 
\nonu
\phi \A = \A \xi_\phi 
- \kappa \epsilon^{ij} \bar\psi^i \gamma_5 \xi_\lambda^j 
- {1 \over 2} \xi \kappa \epsilon^{ij} \bar\psi^i \gamma_5 \psi^j 
- i \kappa \epsilon^{ij} \bar\psi^i \gamma_5 \!\!\not\!\partial \Lambda^j(\psi) 
+ {1 \over 2} \kappa^2 \epsilon^{ij} \bar\psi^i \gamma_5 \psi^j \Box C(\psi), 
\nonu
v^a \A = \A \xi_v^a 
- i \kappa \epsilon^{ij} \bar\psi^i \gamma^a \xi_\lambda^j 
- {i \over 2} \xi \kappa \epsilon^{ij} \bar\psi^i \gamma^a \psi^j 
- \kappa \epsilon^{ij} \bar\psi^i \!\!\not\!\partial \gamma^a \Lambda^j(\psi) 
+ {i \over 2} \kappa^2 \epsilon^{ij} \bar\psi^i \gamma^a \psi^j \Box C(\psi) 
\nonu
\A \A 
- i \kappa^2 \epsilon^{ij} \bar\psi^i \gamma^b \psi^j \partial^a \partial_b C(\psi), 
\nonu
\lambda^i \A = \A \xi_\Lambda^i 
+ \xi \psi^i - i \kappa \!\!\not\!\partial M^{ij}(\psi) \psi^j 
+ {i \over 2} \kappa \epsilon^{ab} \epsilon^{ij} \gamma_a \psi^j \partial_b \phi(\psi) 
\nonu
\A \A 
- {1 \over 2} \kappa \epsilon^{ij} \left\{ \psi^j \partial_a v^a(\psi) 
- {1 \over 2} \epsilon^{ab} \gamma_5 \psi^j F_{ab}(\psi) \right\} 
\nonu
\A \A
- {1 \over 2} \kappa^2 \{ \Box \Lambda^i(\psi) \bar\psi^j \psi^j - \Box \Lambda^j(\psi) \bar\psi^i \psi^j 
- \gamma_5 \Box \Lambda^j(\psi) \bar\psi^i \gamma_5 \psi^j 
\nonu
\A \A 
- \gamma_a \Box \Lambda^j(\psi) \bar\psi^i \gamma^a \psi^j 
+ 2 \!\!\not\!\partial \partial_a \Lambda^j(\psi) \bar\psi^i \gamma^a \psi^j \} 
- {i \over 2} \kappa^3 \!\!\not\!\partial \Box C(\psi) \psi^i \bar\psi^j \psi^j, 
\nonu
D \A = \A {\xi \over \kappa} - i \kappa \bar\psi^i \!\!\not\!\partial \lambda^i(\psi) 
\nonu
\A \A 
+ {1 \over 2} \kappa^2 \left\{ \bar\psi^i \psi^j \Box M^{ij}(\psi) 
- {1 \over 2} \epsilon^{ij} \bar\psi^i \gamma_5 \psi^j \Box \phi(\psi) \right. 
\nonu
\A \A 
\left. 
+ {i \over 2} \epsilon^{ij} \bar\psi^i \gamma_a \psi^j \Box v^a(\psi) 
- i \epsilon^{ij} \bar\psi^i \gamma_a \psi^j \partial_a \partial_b v^b(\psi) \right\} 
\nonu
\A \A
- {i \over 2} \kappa^3 \bar\psi^i \psi^i \bar\psi^j \!\!\not\!\partial \Box \Lambda^j(\psi) 
+ {1 \over 8} \kappa^4 \bar\psi^i \psi^i \bar\psi^j \psi^j \Box^2 C(\psi), 
\label{SUSYrelation-VSF}
\ea
while the SUSY invariant relations $\varphi_\Phi^I = \varphi_\Phi^I(\psi)$ are 
\ba
B^i \A = \A \xi_B^i + \kappa (\bar\psi^i \xi_\chi - \epsilon^{ij} \bar\psi^j \xi_\nu) 
- {1 \over 2} \kappa^2 \{ \bar\psi^j \psi^j F^i(\psi) - 2 \bar\psi^i \psi^j F^j(\psi) 
+ 2 i \bar\psi^i \!\!\not\!\partial B^j(\psi) \psi^j \} 
\nonu
\A \A 
- i \kappa^3 \bar\psi^j \psi^j \{ \bar\psi^i \!\!\not\!\partial \chi(\psi) 
- \epsilon^{ik} \bar\psi^k \!\!\not\!\partial \nu(\psi) \} 
+ {3 \over 8} \kappa^4 \bar\psi^j \psi^j \bar\psi^k \psi^k \Box B^i(\psi), 
\nonu
\chi \A = \A \xi_\chi + \kappa \{ \psi^i F^i(\psi) - i \!\!\not\!\partial B^i(\psi) \psi^i \} 
\nonu
\A \A 
- {i \over 2} \kappa^2 [ \not\!\partial \chi(\psi) \bar\psi^i \psi^i 
- \epsilon^{ij} \{ \psi^i \bar\psi^j \!\!\not\!\partial \nu(\psi) 
- \gamma^a \psi^i \bar\psi^j \partial_a \nu(\psi) \} ] 
\nonu
\A \A 
+ {1 \over 2} \kappa^3 \psi^i \bar\psi^j \psi^j \Box B^i(\psi) 
+ {i \over 2} \kappa^3 \!\!\not\!\partial F^i(\psi) \psi^i \bar\psi^j \psi^j 
+ {1 \over 8} \kappa^4 \Box \chi(\psi) \bar\psi^i \psi^i \bar\psi^j \psi^j, 
\nonu
\nu \A = \A \xi_\nu - \kappa \epsilon^{ij} \{ \psi^i F^j(\psi) - i \!\!\not\!\partial B^i(\psi) \psi^j \} 
\nonu
\A \A 
- {i \over 2} \kappa^2 [ \not\!\partial \nu(\psi) \bar\psi^i \psi^i 
+ \epsilon^{ij} \{ \psi^i \bar\psi^j \!\!\not\!\partial \chi(\psi) 
- \gamma^a \psi^i \bar\psi^j \partial_a \chi(\psi) \} ] 
\nonu
\A \A 
+ {1 \over 2} \kappa^3 \epsilon^{ij} \psi^i \bar\psi^k \psi^k \Box B^j(\psi) 
+ {i \over 2} \kappa^3 \epsilon^{ij} \!\!\not\!\partial F^i(\psi) \psi^j \bar\psi^k \psi^k 
+ {1 \over 8} \kappa^4 \Box \nu(\psi) \bar\psi^i \psi^i \bar\psi^j \psi^j, 
\nonu
F^i \A = \A {\xi^i \over \kappa} - i \kappa \{ \bar\psi^i \!\!\not\!\partial \chi(\psi) 
+ \epsilon^{ij} \bar\psi^j \!\!\not\!\partial \nu(\psi) \} 
\nonu
\A \A 
- {1 \over 2} \kappa^2 \bar\psi^j \psi^j \Box B^i(\psi) + \kappa^2 \bar\psi^i \psi^j \Box B^j(\psi) 
+ i \kappa^2 \bar\psi^i \!\!\not\!\partial F^j(\psi) \psi^j 
\nonu
\A \A 
+ {1 \over 2} \kappa^3 \bar\psi^j \psi^j \{ \bar\psi^i \Box \chi(\psi) + \epsilon^{ik} \bar\psi^k \Box \nu(\psi) \} 
- {1 \over 8} \kappa^4 \bar\psi^j \psi^j \bar\psi^k \psi^k \Box F^i(\psi). 
\label{SUSYrelation-SSF}
\ea

For simplicity of arguments about NL/L SUSY relation, 
we reduce the above SUSY invariant constraints and the SUSY invariant relations 
(the massless eigenstates composed of $\psi^i$) to more simple (but nontrivial and general) expressions. 
Since in Eqs.(\ref{SUSYrelation-VSF}) and (\ref{SUSYrelation-SSF}) 
the constants (the v.e.v.'s) which do not couple to $\psi^i$ are only $\xi_c$ and $\xi_B^i$, we put 
\be
\xi_\Lambda^i = \xi_M^{ij} = \xi_\phi = \xi_v^a = \xi_\lambda^i = 0, 
\ \ \xi_\chi = \xi_\nu = 0. 
\ee
except for $\xi$ and $\xi^i$ which are the fundamental constants in the simplest 
and nontrivial NL/L SUSY relation for the $N = 2$ SUSY QED theory with the SSB \cite{lin-ST3,lin-ST4b}. 
Further we put 
\be
\xi_B^i = 0, 
\ee
because we would like to attribute straightforwardly the $N = 2$ SUSY QED action (\ref{SQEDaction}) 
to the $N = 2$ NLSUSY action (\ref{NLSUSYaction}) up to a normalization factor 
when the SUSY invariant relations are substituted into Eq.(\ref{SQEDaction}). 
Then, the SUSY invariant constraints (\ref{SUSYconst-VSF1}) and (\ref{SUSYconst-SSF1}) become 
\ba
\A \A 
\tilde C = \xi_c, \ \ \tilde\Lambda^i = \tilde M^{ij} = \tilde\phi = \tilde v^a = \tilde\lambda^i = 0, 
\ \ \tilde D = {\xi \over \kappa}, 
\label{SUSYconst-VSF2}
\\
\A \A 
\tilde B^i = \tilde\chi = \tilde\nu = 0, \ \ \ \tilde F^i = {\xi^i \over \kappa}, 
\label{SUSYconst-SSF2}
\ea
and the SUSY invariant relations (\ref{SUSYrelation-VSF}) and (\ref{SUSYrelation-SSF}) reduce to 
\ba
C \A = \A \xi_c - {1 \over 8} \xi \kappa^3 \bar\psi^i \psi^i \bar\psi^j \psi^j \vert w \vert, 
\nonu
\Lambda^i \A = \A - {1 \over 2} \xi \kappa^2 
\psi^i \bar\psi^j \psi^j \vert w \vert, 
\nonu
M^{ij} \A = \A {1 \over 2} \xi \kappa \bar\psi^i \psi^j \vert w \vert, 
\nonu
\phi \A = \A - {1 \over 2} \xi \kappa \epsilon^{ij} \bar\psi^i \gamma_5 \psi^j \vert w \vert, 
\nonu
v^a \A = \A - {i \over 2} \xi \kappa \epsilon^{ij} \bar\psi^i \gamma^a \psi^j \vert w \vert, 
\nonu
\lambda^i \A = \A \xi \psi^i \vert w \vert, 
\nonu
D \A = \A {\xi \over \kappa} \vert w \vert, 
\label{SUSYrelation-VSF1}
\end{eqnarray}
and 
\ba
\chi \A = \A \xi^i \left[ \psi^i \vert w \vert
+ {i \over 2} \kappa^2 \partial_a 
( \gamma^a \psi^i \bar\psi^j \psi^j \vert w \vert 
) \right], 
\nonu
B^i \A = \A - \kappa \left( {1 \over 2} \xi^i \bar\psi^j \psi^j 
- \xi^j \bar\psi^i \psi^j \right) \vert w \vert, 
\nonu
\nu \A = \A \xi^i \epsilon^{ij} \left[ \psi^j \vert w \vert 
+ {i \over 2} \kappa^2 \partial_a 
( \gamma^a \psi^j \bar\psi^k \psi^k \vert w \vert 
) \right], 
\nonu
F^i \A = \A {1 \over \kappa} \xi^i \left\{ \vert w \vert 
+ {1 \over 8} \kappa^3 
\Box ( \bar\psi^j \psi^j \bar\psi^k \psi^k \vert w \vert ) 
\right\} 
\nonu
\A \A 
- i \kappa \xi^j \partial_a ( \bar\psi^i \gamma^a \psi^j \vert w \vert ), 
\label{SUSYrelation-SSF1}
\end{eqnarray}
which are written in the form containing some vanishing terms due to $(\psi^i)^5 \equiv 0$.

\section{NL/L SUSY relation for $N = 2$ SUSY QED}

In this section we discuss the relation between the $N = 2$ SUSY QED action (\ref{SQEDaction}) 
and the $N = 2$ NLSUSY action (\ref{NLSUSYaction}). 
Substituting the reduced (but general) SUSY invariant relations (\ref{SUSYrelation-VSF1}) 
and (\ref{SUSYrelation-SSF1}) into Eqs.(\ref{Vkin}), (\ref{VFI}) and (\ref{gauge}) 
gives the relations among the actions as follows; 
\ba
L_{{\cal V}{\rm kin}}(\psi) \A = \A - \xi^2 L_{N = 2{\rm NLSUSY}}, 
\nonu
L_{{\cal V}{\rm FI}}(\psi) \A = \A 2 \xi^2 L_{N = 2{\rm NLSUSY}}, 
\nonu
(L_{\Phi{\rm kin}} + L_e)(\psi) \A = \A - (\xi^i)^2 e^{-4 e \xi_c} L_{N = 2{\rm NLSUSY}}, 
\label{NL-LSUSY}
\ea
which can be obtained systematically by changing the integration variables 
in the actions (\ref{Vkin}), (\ref{VFI}) and (\ref{gauge}) 
from $(x, \theta^i)$ to $(x', \theta'^i)$ under the SUSY invariant constraints 
(\ref{SUSYconst-VSF2}) and (\ref{SUSYconst-SSF2}) (see, for example, \cite{lin-ST2}). 
Therefore, from Eq.(\ref{NL-LSUSY}) we obtain a general NL/L SUSY relation 
for the $N = 2$ SUSY QED theory in $d = 2$ as 
\be
f(\xi, \xi^i, \xi_c, e) \ L_{N = 2{\rm NLSUSY}} = L^{\rm gen.}_{N = 2{\rm SUSYQED}} 
\label{NL-LSUSYgen}
\ee
with a normalization factor $f(\xi, \xi^i, \xi_c, e)$ defined by 
\be
f(\xi, \xi^i, \xi_c, e) = \xi^2 - (\xi^i)^2 e^{-4 e \xi_c}. 
\label{normalization}
\ee

Here we remind ourselves of the NLSUSY GR model in SGM scenario. 
In the SGM action $L_{\rm SGM}(e,\psi)$ 
the relative scale of the kinetic terms for the NG fermions $\psi^i$ to the Einstein GR theory is fixed as 
$-{1 \over {2 \kappa^2}} t^a{}_a = {i \over 2} \bar\psi^i \!\!\not\!\!\partial \psi^i$; 
namely, the SGM action can be written (in $d = 4$) as \cite{KS1,ST1,ST2} 
\be
L_{\rm SGM}(e,\psi) = {c^4 \over {16 \pi G}} e ( R - \vert w \vert \Lambda + \cdots ), 
\label{SGM}
\ee
and then the dimensional constant $\kappa$ in the determinant (\ref{det-w}) 
is fixed to $\kappa^{-2} = {{c^4 \Lambda} \over {8 \pi G}}$. 
By regarding the NLSUSY GR theory in SGM scenario as the fundamental theory of space-time and matter, 
$L^{\rm gen.}_{N = 2{\rm SUSYQED}}$ is attributed to $L_{N = 2{\rm NLSUSY}}$ which is the cosmological term 
of SGM (NLSUSY GR) action for the asymptotically flat space-time, i.e. we put 
\be
f(\xi, \xi^i, \xi_c, e) = 1. 
\label{normalization1}
\ee
This situation is different from the string theory, where (the magnitude of the coupling contant of) the 
gauge theory appears by the compactification of extra space. 

Remarkably, the condition (\ref{normalization1}) gives the gauge coupling constant $e$ 
in terms of $\xi$, $\xi^i$ and $\xi_c$ as 
\be
e = {1 \over 4\xi_c} \ln X, 
\ \ \ X = {(\xi^i)^2 \over {\xi^2 - 1}}. 
\label{gcoupling}
\ee

Now let us discuss the relation between the general $N = 2$ SUSY QED action (\ref{SQEDaction}) 
and the {\it minimal} one for the minimal off-shell vector supermultiplet 
in the NL/L SUSY relation (\ref{NL-LSUSYgen}) with the normalization condition (\ref{normalization}). 
The (minimal) actions $L^0_{{\cal V}{\rm kin}}$, $L^0_{{\cal V}{\rm FI}}$, 
$L^0_{\Phi{\rm kin}}$ and $L^0_e$ defined in Eqs. from (\ref{Vkin-comp}) to (\ref{gauge-comp}) 
is related to the general $N = 2$ SUSY QED action of Eq.(\ref{NL-LSUSY}) in NL/L SUSY relation as 
\ba
L_{{\cal V}{\rm kin}}(\psi) \A = \A L^0_{{\cal V}{\rm kin}}(\psi) = - \xi^2 L_{N = 2{\rm NLSUSY}}, 
\nonu
L_{{\cal V}{\rm FI}}(\psi) \A = \A L^0_{{\cal V}{\rm FI}}(\psi) + [{\rm tot.\ der.\ terms}] 
= 2 \xi^2 L_{N = 2{\rm NLSUSY}}, 
\nonu
(L_{\Phi{\rm kin}} + L_e)(\psi) 
\A = \A (e^{-4 e \xi_c} L^0_{\Phi{\rm kin}}\vert_{F \rightarrow F'} + L^0_e)(\psi) 
+ [{\rm tot.\ der.\ terms}] 
\nonu
\A = \A - (\xi^i)^2 e^{-4 e \xi_c} L_{N = 2{\rm NLSUSY}}, 
\label{NL-LSUSY0}
\ea
where $L^0_e(\psi) = {1 \over 4} e \kappa \xi (\xi^i)^2 \bar\psi^j \psi^j\bar\psi^k \psi^k$ 
and the SUSY invariant relations of the auxiliary fields $F^i$ in Eq.(\ref{SUSYrelation-SSF1}) 
have been changed (relaxed) by four NG fermion self-interaction terms as 
\be
F'^i(\psi) = F^i(\psi) - {1 \over 4} e^{4 e \xi_c} e \kappa^2 \xi \xi^i \bar\psi^j \psi^j \bar\psi^k \psi^k. 
\ee
Obviously, the minimal $N = 2$ SUSY QED action for the minimal off-shell vector supermultiplet 
is included in the relations (\ref{NL-LSUSY0}) at the leading order of the factor $e^{-4 e \xi_c}$. 

It can be seen easily that the numerical factor $e^{-4 e \xi_c}$ in the relation (\ref{NL-LSUSY0}) 
is absorbed into the action by rescaling the whole scalar supermultiplet $\Phi^i$ by $e^{-2 e \xi_c}$ 
and by translating the auxiliary field $C$ by $\xi_c$ in the gauge action (\ref{gauge}) as 
\ba
\A \A 
\varphi_\Phi^I(\psi) \rightarrow \hat \varphi_\Phi^I(\psi) = e^{-2 \xi_C e} \varphi_\Phi^I(\psi), 
\nonu
\A \A 
C(\psi) \rightarrow \hat C(\psi) = - {1 \over 8} \xi \kappa^3 \bar\psi^i \psi^i \bar\psi^j \psi^j \vert w \vert. 
\label{rescale}
\ea
Indeed, the actions (\ref{Vkin}), (\ref{VFI}) and (\ref{gauge}) 
in terms of the fields $\{ \hat \varphi_\Phi^I, \hat C \}$ become 
\ba
L_{{\cal V}{\rm kin}}(\psi) \A = \A \hat L_{{\cal V}{\rm kin}}(\psi) 
= \hat L^0_{{\cal V}{\rm kin}}(\psi) = - \xi^2 L_{N = 2{\rm NLSUSY}}, 
\label{Vkin-hat}
\\
L_{{\cal V}{\rm FI}}(\psi) \A = \A \hat L_{{\cal V}{\rm FI}}(\psi) 
= \hat L^0_{{\cal V}{\rm FI}}(\psi) + [{\rm tot.\ der.\ terms}] 
\nonu
\A = \A 2 \xi^2 L_{N = 2{\rm NLSUSY}}, 
\label{VFI-hat}
\\
(L_{\Phi{\rm kin}} + L_e)(\psi) \A = \A (\hat L_{\Phi{\rm kin}} + \hat L_e)(\psi) 
\nonu
\A = \A (\hat L^0_{\Phi{\rm kin}}\vert_{\hat F \rightarrow \hat F'} + \hat L^0_e)(\psi) 
+ [{\rm tot.\ der.\ terms}] 
\nonu
\A = \A - (\xi^i)^2 e^{-4 e \xi_c} L_{N = 2{\rm NLSUSY}}, 
\label{gauge-hat}
\ea
where 

\be
\hat F'^i(\psi) =  e^{-2 e \xi_c} \left\{ F^i(\psi) 
- {1 \over 4}e \kappa^2 \xi \xi^i \bar\psi^j \psi^j \bar\psi^k \psi^k \right\}. 
\ee

Therefore, we obtain the ordinary $N = 2$ SUSY QED action for the minimal off-shell vector supermultiplet 
with the $U(1)$ gauge coupling constant (\ref{gcoupling}), i.e. 
\be
L_{N = 2{\rm NLSUSY}} = L^{\rm gen.}_{N = 2{\rm SUSYQED}} 
= \hat L^0_{N = 2{\rm LSUSYQED}} + [{\rm tot.\ der.\ terms}], 
\label{NL-LSUSYgen0}
\ee
where the minimal $N = 2$ SUSY QED ($U(1)$ gauge) action $\hat L^0_{N = 2{\rm LSUSYQED}}$ is defined by 
\be
\hat L^0_{N = 2{\rm LSUSYQED}} 
= \hat L^0_{{\cal V}{\rm kin}} + \hat L^0_{{\cal V}{\rm FI}} 
+ \hat L^0_{\Phi{\rm kin}}\vert_{\hat F \rightarrow \hat F'} + \hat L^0_e. 
\label{minSQEDaction}
\ee
Interestingly $e$ defined by the action (\ref{gauge}) depends upon the constant terms (the v.e.v.'s) 
of the auxiliary fields, i.e. the vacuum structures. 
(Note that the bare $e$ is an arbitrary parameter, provided  $\xi_c \rightarrow 0$ corresponding to 
the adoption of the WZ gauge throughout the arguments.)

\section{Summary and discussions}

In this paper we have studied the NL/L SUSY relation for the $N = 2$ SUSY QED theory in $d = 2$ 
starting from the most general SUSY invariant constraints (\ref{SUSYconst-VSF1}) and (\ref{SUSYconst-SSF1}) 
and the subsequent SUSY invariant relations (\ref{SUSYrelation-VSF}) and (\ref{SUSYrelation-SSF}). 
After reducing those constraints and relations to simpler (but general) expressions 
of Eqs. from (\ref{SUSYconst-VSF2}) to (\ref{SUSYrelation-SSF1}), 
we have obtained the general NL/L SUSY invariant relation (\ref{NL-LSUSYgen}) 
which produces the overall (normalization) factor (\ref{normalization}) depending 
on the gauge coupling constant $e$. 
The overall factor (compositeness condition for all particles) should be (\ref{normalization1})  
in SGM scenario for the basic NLSUSY GR theory, which gives the relations 
between $e$ and the constant terms in SUSY invariant relations (the v.e.v.'s) of the auxiliary fields, 
$\xi$, $\xi^i$ and $\xi_c$ as in Eq.(\ref{gcoupling}). 
We have shown  in detail that Eq.(\ref{gcoupling}) holds in the subsequent WZ gauge as well.  

Our study may indicate that the general structure of the constant terms (the v.e.v.'s) of auxiliary fields 
for the general (gauge) superfield and the NL/L SUSY relations 
play  crucial roles in SUSY (composite) theory by determining the true vacuum of NLSUSY GR (SGM) 
through the Big Decay and the SSB,  
explaining simply the observed mysterious numerical relations 
between {\it the (dark) energy density of the universe} $\rho_D$ ($\sim {{c^4 \Lambda} \over {8 \pi G}}$) 
at the large scale  and {\it the neutrino mass} at the (low energy) local frame $m_\nu$ \cite{ST5,STL}, 
\be
\rho_D^{\rm obs} \sim (10^{-(12 \sim 13)} GeV)^4 \sim (m_\nu){}^4 
\sim {\Lambda \over G} \ (\sim {g_{sv}}^2), 
\label{darkrelation}
\ee
(where $g_{sv}$ is the coupling constant of superon with the vacuum)
and determining the magnitude of the (bare) gauge coupling constant 
(i.e. the SUSY compositeness condition for all particles including the auxiliary fields).  
These are characteristic features of the SGM scenario for unity of space-time and matter, where  
the ordinaly LSUSY theory in flat space-time  originates from the cosmological term of 
NLSUSY GR action. 
The similar arguments in $d = 4$ for more general SUSY invariant constraints 
and for the large $N$ SUSY, especially $N = 4, 5$ are interesting and crucial. 

Finally we just mention that Eq.(\ref{darkrelation}) of SGM scenario may indicate 
the superfluidity of nature, i.e. what Landau-Gintzburg theory is to BCS theory of superconductivity, 
LSUSY theory is to NLSUSY GR (SGM) of the superfluidity of space-time and matter. 
The observed particles of LSUSY theory may be the SUSY composites (eigenstates) of NG fermions 
dictated by the space-time symmetry.

\newpage

%
\newcommand{\NP}[1]{{\it Nucl.\ Phys.\ }{\bf #1}}
\newcommand{\PL}[1]{{\it Phys.\ Lett.\ }{\bf #1}}
\newcommand{\CMP}[1]{{\it Commun.\ Math.\ Phys.\ }{\bf #1}}
\newcommand{\MPL}[1]{{\it Mod.\ Phys.\ Lett.\ }{\bf #1}}
\newcommand{\IJMP}[1]{{\it Int.\ J. Mod.\ Phys.\ }{\bf #1}}
\newcommand{\PR}[1]{{\it Phys.\ Rev.\ }{\bf #1}}
\newcommand{\PRL}[1]{{\it Phys.\ Rev.\ Lett.\ }{\bf #1}}
\newcommand{\PTP}[1]{{\it Prog.\ Theor.\ Phys.\ }{\bf #1}}
\newcommand{\PTPS}[1]{{\it Prog.\ Theor.\ Phys.\ Suppl.\ }{\bf #1}}
\newcommand{\AP}[1]{{\it Ann.\ Phys.\ }{\bf #1}}


\begin{thebibliography}{100}
\bibitem{WZ}
J. Wess and B. Zumino, {\it Phys. Lett. B} {\bf 49} (1974) 52. 

\bibitem{VAa}
D.V. Volkov and V.P. Akulov, {\it Phys. Lett. B} {\bf 46} (1973) 109. 

\bibitem{KS1}
K. Shima, {\it Phys. Lett. B} {\bf 501} (2001) 237. 

\bibitem{VS}
D.V. Volkov and V.A. Soroka,  
{\it JETP Lett.} {\bf 18} (1973) 312. 

\bibitem{ST1}
K. Shima and M. Tsuda, {\it Phys. Lett. B} {\bf 507} (2001) 260. 

\bibitem{ST2}
K. Shima and M. Tsuda, {\it PoS HEP2005} (2006) 011. 

\bibitem{KS2}
K. Shima, {\it European Phys. J. C} {\bf 7} (1999) 341. 

\bibitem{IK1}
E.A. Ivanov and A.A. Kapustnikov, {\it J. Phys. A} {\bf 11} (1978) 2375. 

\bibitem{Ro}
M. Ro\v{c}ek, {\it Phys. Rev. Lett.} {\bf 41} (1978) 451. 

\bibitem{UZ}
T. Uematsu and C.K. Zachos, {\it Nucl. Phys. B} {\bf 201} (1982) 250. 

\bibitem{STT1}
K. Shima, Y. Tanii and M. Tsuda, {\it Phys. Lett. B} {\bf 525} (2002) 183. 

\bibitem{STT2}
K. Shima, Y. Tanii and M. Tsuda, {\it Phys. Lett. B} {\bf 546} (2002) 162. 

\bibitem{ST3}
K. Shima and M. Tsuda, {\it Phys. Lett. B} {\bf 641} (2006) 101. 

\bibitem{lin-ST2}
K. Shima and M. Tsuda, {\it Mod. Phys. Lett. A} {\bf 23} (2008) 3149. 

\bibitem{lin-ST3}
K. Shima and M. Tsuda, {\it Mod. Phys. Lett. A} {\bf 22} (2007) 3027.

\bibitem{lin-ST4a}
K. Shima and M. Tsuda, {\it Phys. Lett. B} {\bf 666} (2008) 410. 

\bibitem{lin-ST4b}
K. Shima and M. Tsuda, {\it Mod. Phys. Lett. A} {\bf 24} (2009) 185. 

\bibitem{WB}
J. Wess and J. Bagger, {\it Supersymmetry and Supergravity (Second Edition)} 
(Princeton University Press, Princeton, New Jersey, 1992). 

\bibitem{ST5}
K. Shima and M. Tsuda, {\it Phys. Lett. B} {\bf 645} (2007) 455. 

\bibitem{STL}
K. Shima, M. Tsuda and W. Lang, {\it Phys. Lett. B} {\bf 659} (2008) 741. 

\bibitem{DVF}
P. Di Vecchia and S. Ferrara, \NP{B130} (1977) 93. 

\bibitem{ST4}
K. Shima and M. Tsuda, \MPL{A23} (2008) 1167. 




















\end{thebibliography}
\end{document}